\documentclass[10pt, conference, compsocconf]{IEEEtran}

\usepackage[american]{babel}
\usepackage{cite}
\usepackage[T1]{fontenc}
\usepackage{tikz}
\usepackage{amsmath}
\usepackage{hhline}
\usepackage{comment}
\usepackage[textsize=scriptsize]{todonotes}
\usepackage{pifont}
\let\labelindent\relax
\usepackage{enumitem}
\usepackage{graphicx}
\usepackage{subcaption}
\usepackage{soul}
\usepackage{xspace}
\usepackage[hyphens]{url}
\usepackage{hyperref}
\usepackage[hyphenbreaks]{breakurl}
\usepackage[keeplastbox]{flushend} %
\usepackage{booktabs}
\usepackage{url}

\usepackage[toc,acronym,nowarn]{glossaries}

\newacronym{OS}{OS}{Operating System}
\newacronym{SE}{SE}{Secure Element}
\newacronym{NFC}{NFC}{Near Field Communication}
\newacronym{BLE}{BLE}{Bluetooth Low Energy}
\newacronym{RIL}{RIL}{Radio Interface Layer}
\newacronym{SELinux}{SELinux}{Security-Enhanced Linux}
\newacronym{LSM}{LSM}{Linux Security Modules}
\newacronym{TCB}{TCB}{Trusted Computing Base}
\newacronym{MAC}{MAC}{Mandatory Access Control}
\newacronym{IPC}{IPC}{Inter-Process Communication}
\newacronym{ICC}{ICC}{Inter-Container Communication}
\newacronym{ISA}{ISA}{Instruction Set Architecture}
\newacronym{cgroups}{cgroups}{control groups}
\newacronym{CA}{CA}{Certificate Authority}
\newacronym{PKI}{PKI}{Public Key Infrastructure}
\newacronym{MITM}{MITM}{Man-In-The-Middle}
\newacronym{BYOD}{BYOD}{Bring-Your-Own-Device}
\newacronym{CM}{CM}{Container Management}
\newacronym{SM}{SM}{Security Management}
\newacronym{HAL}{HAL}{Hardware Abstraction Layer}
\newacronym{TLS}{TLS}{Transport Layer Security}
\newacronym{C2C}{C2C}{Container To Container}
\newacronym{Protobuf}{Protobuf}{Protocol Buffers}
\newacronym{SDO}{SDO}{Sensitive Data Object}
\newacronym{VMA}{VMA}{Virtual Memory Area}
\newacronym{PGD}{PGD}{Page Global Directory}
\newacronym{PUD}{PUD}{Page Upper Directory}
\newacronym{PMD}{PMD}{Page Middle Directory}
\newacronym[firstplural=Page Table Entries (PTEs)]{PTE}{PTE}{Page Table Entry}
\newacronym{COW}{COW}{Copy-On-Write}
\newacronym{IV}{IV}{Initialization Vector}
\newacronym{ESSIV}{ESSIV}{Encrypted Salt-Sector Initialization Vector}
\newacronym{KSM}{KSM}{Kernel Samepage Merging}
\newacronym{JIT}{JIT}{Just-In-Time}
\newacronym{DMA}{DMA}{Direct Memory Access}
\newacronym{FDE}{FDE}{Full Disk Encryption}
\newacronym{AS}{AS}{Address Space}
\newacronym{GCM}{GCM}{Google Cloud Messaging}
\newacronym{TPM}{TPM}{Trusted Platform Module}
\newacronym{JTAG}{JTAG}{Joint Test Action Group}
\newacronym{LUKS}{LUKS}{Linux Unified Key Setup}
\newacronym{VPN}{VPN}{Virtual Private Network}
\newacronym{PBKDF2}{PBKDF2}{Password-Based Key Derivation Function 2}
\newacronym{VM}{VM}{Virtual Machine}
\newacronym{HV}{HV}{Hypervisor}
\newacronym{SEV}{SEV}{Secure Encrypted Virtualization}
\newacronym{SME}{SME}{Secure Memory Encryption}
\newacronym{TSME}{TSME}{Transparent \gls{SME}}
\newacronym{SP}{SP}{Secure Processor}
\newacronym{GPT}{GPT}{Guest Page Table}
\newacronym{HPT}{HPT}{Host Page Table}
\newacronym{TLB}{TLB}{Translation Lookaside Buffer}
\newacronym{PoC}{PoC}{Proof of Concept}
\newacronym{ORAM}{ORAM}{Oblivious RAM}
\newacronym{SEV-ES}{SEV-ES}{SEV Encrypted State}
\newacronym{SEV-SNP}{SEV-SNP}{SEV Secure Nested Paging}
\newacronym{RMP}{RMP}{Reverse Map Table}
\newacronym{VMCB}{VMCB}{Virtual Machine Control Block}
\newacronym{SLAT}{SLAT}{Second Level Address Translation}
\newacronym{SSH}{SSH}{Secure Shell}
\newacronym{RSA}{RSA}{Rivest–Shamir–Adleman}
\newacronym{ECDHE}{ECDHE}{Elliptic-Curve Diffie-Hellman Ephemeral}
\newacronym{AES}{AES}{Advanced Encryption Standard}
\newacronym{OOM}{OOM}{Out Of Memory}
\newacronym{MKTME}{MKTME}{Multi-Key Total Memory Encryption}
\newacronym{VMI}{VMI}{Virtual Machine Introspection}
\newacronym{MAD}{MAD}{Median Absolute Deviation}
\newacronym{HSM}{HSM}{Hardware Security Module}
\newacronym{AE}{AE}{Automatic Exit}
\newacronym{NAE}{NAE}{Non-Automatic Exit}
\newacronym{AES-NI}{AES-NI}{AES New Instructions}
\newacronym{NIC}{NIC}{Network Interface Card}
\newacronym{NMI}{NMI}{Non-Maskable Interrupt}
\newacronym{MTU}{MTU}{Maximum Transmission Unit}
\newacronym{VA}{VA}{Virtual Address}
\newacronym{GFN}{GFN}{Guest Frame Number}
\newacronym{SFN}{SFN}{System Frame Number}
\newacronym{IOMMU}{IOMMU}{I/O Memory Management Unit}
\newacronym{AISE}{AISE}{Address Independent Seed Encryption}
\newacronym{MT}{MT}{Merkle Tree}
\newacronym{BMT}{BMT}{Bonsai Merkle Tree}
\newacronym{LPID}{LPID}{Located Page IDentifier}
\newacronym{CB}{CB}{Counter Block}
\newacronym{PRD}{PRD}{Page Root Directory}
\newacronym{SWIOTLB}{SWIOTLB}{Software I/O Translation Buffer}
\newacronym{ASID}{ASID}{Address Space Identifier}
\newacronym{vCPU}{vCPU}{virtual CPU}
\newacronym{VC}{\texttt{\#VC}}{VMM Communication Exception}
\newacronym{GHCB}{GHCB}{Guest Hypervisor Communication Block}
\newacronym{IDT}{IDT}{Interrupt Descriptor Table}
\newacronym{KASLR}{KASLR}{Kernel Address Space Layout Randomization}
\newacronym{SLES}{SLES}{SUSE Linux Enterprise Server}
\newacronym{RHEL}{RHEL}{RedHat Enterprise Linux}
\newacronym{IBS}{IBS}{Instruction Based Sampling}
\newacronym{CFM}{CFM}{Control Flow Modification}
\newacronym{CE}{CE}{Code Execution}
\newacronym{TSC}{TSC}{Time Stamp Counter}

\makeglossaries
\glsdisablehyper
 
\graphicspath{{figures/}}

\addto\extrasamerican{

}

\newcommand{\phase}[1]{\ding{\numexpr181 + #1}}

\newcommand{\name}{SEVerity\xspace}

\newcommand{\circled}[1]{%
	\tikz[baseline=(char.base)]\node[anchor=north, draw,circle, inner sep=1pt,fill=black,text=white, scale=0.9, minimum size=12pt](char){#1} ;}

\begin{document}

\title{SEVerity: Code Injection Attacks against Encrypted~Virtual~Machines}

    \author{
      \IEEEauthorblockN{
        Mathias Morbitzer\IEEEauthorrefmark{1},
        Sergej Proskurin\IEEEauthorrefmark{2},
        Martin Radev\IEEEauthorrefmark{1},
        Marko Dorfhuber\IEEEauthorrefmark{2}
        and Erick Quintanar Salas\IEEEauthorrefmark{1}
    }
    \IEEEauthorblockA{\IEEEauthorrefmark{1}Fraunhofer AISEC\\
    \{mathias.morbitzer,martin.radev,erick.quintanar.salas\}@aisec.fraunhofer.de}\\
    \IEEEauthorblockA{\IEEEauthorrefmark{2}Technical University of Munich\\
    proskurin@sec.in.tum.de, marko.dorfhuber@tum.de}
}

\maketitle

\begin{abstract}
Modern enterprises increasingly take advantage of cloud infrastructures. 
Yet, outsourcing code and data into the cloud requires enterprises to trust cloud providers not to meddle with their data. 
To reduce the level of trust towards cloud providers, AMD has introduced 
\gls{SEV}.
By encrypting \glspl{VM}, 
\gls{SEV} aims to ensure %
data confidentiality,
despite a compromised or curious Hypervisor.
The \gls{SEV-ES} extension additionally protects the \gls{VM}'s register state from unauthorized access. 
Yet, both extensions do not provide
integrity of the \gls{VM}'s memory, which has already been abused 
to leak the protected data or to alter the \gls{VM}'s control-flow.

In this paper, we introduce the \emph{\name} attack; %
a missing puzzle
piece in the series of attacks against the AMD \gls{SEV} %
family.
Specifically, we abuse the system's lack of 
memory
integrity protection to
inject and execute arbitrary code within \gls{SEV-ES}-protected \glspl{VM}.
Contrary to previous 
code execution attacks 
against the AMD \gls{SEV} family,
\name neither relies on a specific CPU version nor on any code gadgets
inside the \gls{VM}.
Instead, \name abuses the fact that 
\gls{SEV-ES} prohibits
direct memory access into the encrypted memory. 
Specifically, \name injects arbitrary code into the encrypted \gls{VM} through 
I/O channels 
and uses the Hypervisor %
to locate and trigger the execution of the encrypted payload.
This allows us to sidestep the 
protection mechanisms
of \gls{SEV-ES}.
Overall, our results demonstrate a success rate of 100\% and hence highlight
that memory integrity protection is an obligation when encrypting \glspl{VM}.
Consequently, our work presents the final stroke in a series of attacks against AMD \gls{SEV} and \gls{SEV-ES} and 
renders the present implementation as incapable of protecting against a curious, vulnerable, or malicious Hypervisor.

\end{abstract}

\section{Introduction}
\label{sec:intro}
\glsresetall

Cloud computing has become omnipresent and %
continues to get increasingly popular~\cite{denisco2019cloud}.
Despite its many advantages, in particular enterprises are reluctant towards entrusting data to provider-controlled infrastructures~\cite{amigorena2019why}. 
This lack of trust is twofold as the valuable data 
hinges on the integrity of the virtualization software, called the \gls{HV}, as well as the provider's trustworthiness.  
In this regard, the \gls{HV}'s integrity can be undermined by adversaries, despite the added isolation and security guarantees of the virtualization
technology~\cite{liu2015thwarting,chen:2017:privwatcher,hua2018epti,xmp_proskurin:2020}.
For instance, in the presence of vulnerabilities, skillful adversaries manage to escape their \glspl{VM} to take control over the \gls{HV} and all \glspl{VM} that share the same system~\cite{vmware2017vmsa, ssd2018oracle, citrix2019hypervisor, citrix2020hypervisor}.
Similarly, attackers with physical access to the machine can run a series of attacks that %
 reveal the \gls{VM}'s memory contents~\cite{becher2005firewire, boileau2006hit, halderman2008lest}.
Besides, in modern cloud infrastructures 
the cloud provider 
is able to
expose potentially sensitive information 
~\cite{lengyel:2014,proskurin:2018a,proskurin:2018b}. 
As such, enterprises have 
to trust the cloud provider
to preserve data integrity and confidentiality~\cite{meixner2012trust}.

To counteract these concerns, AMD has announced 
\gls{SEV}~\cite{kaplan2016amd}. 
\gls{SEV} lends \glspl{VM} the ability to encrypt their memory using a key maintained by a dedicated \emph{\gls{SP}}. %
Consequently, without knowing the respective secret key, neither the attacker
performing cold boot attacks nor the \gls{HV} will be able to reveal the \gls{VM}'s memory contents.
To also reduce information leakage through the \glspl{VM}'s register state, AMD
introduced \emph{\gls{SEV-ES}}~\cite{sev-es}. Unfortunately, it remains
susceptible to attacks which do not rely on the \gls{VM}'s registers.
Another extension, \emph{\gls{SEV-SNP}}, aims to provide memory integrity protection. 
However, \gls{SEV-SNP} is not yet available in hardware. 
Consequently, the missing memory integrity protection in \gls{SEV-ES} continues to jeopardize sensitive data inside encrypted \glspl{VM}.

In this paper, we present \emph{\name}, a new attack against \gls{SEV-ES}.
\name allows a malicious \gls{HV} to 
inject \emph{truly arbitrary} code into the \gls{VM}'s encrypted memory and
cause the \gls{VM} to execute the injected payload.
Contrary to previous work, \name does not limit itself to software
bugs~\cite{radev2020abusing} or any %
memory encryption algorithm that is specific to a CPU
version~\cite{du2017secure,li2019exploiting,wilke2020sevurity}.
Further, \name does not require any out-of-band collected knowledge of in-\gls{VM} applications
~\cite{hetzelt2017security, du2017secure, buhren2018detectability, morbitzer2018severed, werner2019severest, morbitzer2019extracting, li2019exploiting}.
In fact, our attack does not rely on any in-\gls{VM} invariants, except for exported kernel symbols.
Instead, it abuses the I/O channel to inject and execute truly arbitrary payloads in the \gls{VM}'s encrypted memory.

To mount the \name attack, first, we identify the \gls{VM}'s page frame %
holding code that we can trigger from the outside of the \gls{VM}, our \emph{trigger page}. 
Next, we send a network packet with our payload to the \gls{VM}.
Since \gls{SEV-ES} cannot directly pass packets into the encrypted
memory, we use virtio buffers to
track our payload in the \gls{VM}'s memory. %
Finally, we abuse the lack of \gls{SLAT} integrity protection by
remapping the trigger page to the page holding the injected payload, and trigger its execution.

In summary, we make the following contributions:
\begin{itemize}[noitemsep, itemsep=0.0pt, topsep=0.5pt]
   \item We present \name, the first attack against AMD \gls{SEV-ES} 
   which allows %
       to inject and execute arbitrary code in an encrypted \gls{VM} with kernel privileges, %
		with no currently available countermeasures.
    \item We abuse \gls{SEV-ES}' lack of \gls{SLAT} integrity protection %
        and its inability
        to grant DMA to encrypted memory 
        to run \name,
        without relying on flaws of the encryption algorithm's tweak
        function or firmware vulnerabilities.
    \item We evaluate \name on all Linux distributions that are officially
        supported by \gls{SEV-ES} and manage to achieve an
        attack success rate of 100\,\%.
\end{itemize}

Note that we have responsibly disclosed an early version of this paper to AMD.
AMD has documented the attack vector under CVE-2020-12967.

\section{Background}
\label{sec:background}

Full disk encryption has 
gained popularity.
Yet, even on-disk-encrypted data gets exposed when loaded into main
memory~\cite{tang2012cleanos}.
As such, sensitive information, including
passwords, cryptographic keys, or health related data
can become target to attacks against the main memory~\cite{halderman2008lest,
weinmann2012baseband}.

In this section, we outline AMD's hardware extensions for main memory
encryption
introduced in response to the depicted threat.  We focus
on AMD's efforts to encrypt the \gls{VM}'s memory and highlight the
associated limitations.
Particularly, we amplify the lack of \gls{DMA} to
the encrypted memory, which is key for 
\name.

\glsreset{SEV}
\glsreset{SEV-ES}
\glsreset{SME}

\subsection{AMD Memory Encryption}
\label{sec:background:memenc}

\glsreset{SP}

\noindent
\textbf{AMD SME:}
To thwart attacks against the system's main memory,
AMD has introduced \gls{SME}~\cite{kaplan2016amd}, a hardware extension for both full and partial memory encryption.
This 
extension links the system's page tables with its encryption engine, 
allowing
to regulate which memory regions 
to 
encrypted.
Specifically, AMD dedicates the $47$th bit, the \texttt{C-bit}, of the system's \glspl{PTE} to
determine whether the associated page should be encrypted.
A dedicated ARM-based processor, the \gls{SP}, takes over the management of the secret key. %
The secret key 
used by the AES engine is generated on each system reset and 
solely known to the \gls{SP}.  
By setting the \texttt{C-bit} of a particular \gls{PTE}, the 
\gls{SP} encrypts the respective page via a specially tweaked, high performance AES implementation~\cite{kaplan2016amd}.
The tweak incorporates the data's physical address into the encryption to prevent attackers from moving cipher-text blocks.
Yet, \gls{SME} has drawbacks when deployed in cloud environments. 
For one, the memory of \glspl{VM} and the underlying \gls{HV} is encrypted with the same key. 
Thus, curious administrators can access the sensitive data of \glspl{VM}, despite full memory encryption. 
Additionally, having 
gained access to the encrypted memory of one \gls{VM}, a local attacker will be able to similarly access the memory of all remaining \glspl{VM} running on the same system.

\smallskip \noindent
\textbf{AMD SEV:}
AMD \gls{SEV} 
draws upon 
\gls{SME} 
~\cite{kaplan2016amd}. 
Contrary to \gls{SME}, 
\gls{SEV} establishes 
additional isolation by assigning one secret key per isolation domain, including different \glspl{VM} and the \gls{HV} itself.
Trying to access encrypted memory from a different 
domain associated with a different key returns garbled information.  
As such, the physical memory of isolation domains can only be decrypted with the secret key of the domain to which the memory has been assigned. %
By setting the \texttt{C-bit} of \glspl{PTE} inside the \gls{VM}'s page tables, every \gls{VM} can individually control which page 
to encrypt. 
If the \gls{VM} sets the \texttt{C-bit} for a page, its contents will be encrypted with the \gls{VM}'s key.
Only the associated \gls{VM} can access the contents of such pages, which are 
referred to as \emph{private} pages.
By omitting the \texttt{C-bit}, the \gls{VM} defines pages which can be shared with the \gls{HV}.
Depending on the \texttt{C-bit} in the \gls{HV}'s page tables, such \emph{shared} pages can either be encrypted with the \gls{HV}'s key, or remain unencrypted.
In both cases, the \gls{VM} 
and the \gls{HV} are able to access shared pages.

As the secret keys are governed by the \gls{SP} and are inaccessible to the \gls{HV} and \glspl{VM}, there is no way to leak sensitive information 
maintained in 
different isolation domains. 
Yet, potentially sensitive data can still leak through register contents to the underlying \gls{HV}~\cite{werner2019severest}.

\smallskip \noindent
\textbf{AMD SEV-ES:}
In 2017, AMD has announced an iterative extension of the \gls{SEV}, %
\gls{SEV-ES}~\cite{sev-es}, to prevent \glspl{HV} from snooping or modifying the \gls{VM}'s register contents.  
\gls{SEV-ES} additionally lends \glspl{VM} the capability to encrypt and integrity-protect selected register state 
before handing
over control to the \gls{HV},
 rendering the 
 state inaccessible to the \gls{HV}.

Even though \gls{SEV-ES} overall manages to reduce the attack surface of \glspl{VM}, its architecture neglects critical components 
prone to attacks. 
For instance, similar to \gls{SEV}, \gls{SEV-ES} does not protect the \gls{SLAT},
which translates
\glspl{GFN} to \glspl{SFN}~\cite{hetzelt2017security, morbitzer2018severed}. 
Also, \gls{SEV-ES} denies \gls{DMA} to the \gls{VM}'s memory, which must be established through alternative I/O channels~\cite{kaplan2016amd}. 
It is precisely these channels which facilitate a severe attack vector that allows adversaries to inject and execute arbitrary code inside an encrypted \gls{VM}, the analysis and disclosure of which is the main focus of our work.

Note that our work focuses on \mbox{\gls{SEV-ES}}---the latest
available version of the \gls{SEV} family; \name similarly
applies to \gls{SEV}, which faces the same fundamental security issues.

\subsection{DMA into Encrypted Virtual Machines}
\label{sec:background:dma}

Modern \gls{IOMMU} architectures increasingly focus on virtual environments. 
One task of \glspl{IOMMU} is to transparently isolate direct accesses between devices and different \glspl{VM}~\cite{iommu_ben:2006}.  
Unfortunately, this is not true for \gls{SEV-ES}, as it
prohibits \gls{DMA} into \gls{VM}-encrypted memory~\cite{kaplan2016amd}. 
To bypass these limitations, \gls{SEV-ES} makes use of a software \gls{IOMMU}.

Generally, Linux implements an abstract \gls{DMA} interface, hiding details of the low-level implementation. 
This 
grants Linux the flexibility to fall back to the software implementation of the \gls{IOMMU} functionality, 
referred to as \gls{SWIOTLB}.
\gls{SWIOTLB} allows to emulate the \gls{DMA} interface on systems which do not implement a physical \gls{IOMMU}. 
Internally, \gls{SWIOTLB} implements a \emph{bounce buffering} mechanism to 
enable legacy devices to overcome the $4$~GB boundary by 
granting the device access to 
the \emph{bounce buffer}. 
Once the device has written to the bounce buffer, the system 
copies the contents to the buffer maintained by the device driver. 

To overcome the limitations of \gls{SEV-ES} regarding \gls{DMA}, \glspl{VM} use a similar approach. 
On read requests on non-protected \glspl{VM}, the \gls{HV} writes data directly into a buffer in the \gls{VM}'s memory from which the \gls{VM}'s device driver consumes it. 
Yet, \gls{SEV-ES} 
prevents the \gls{HV} from reading or writing meaningful data into the \gls{VM}'s memory. 
Instead, the \gls{HV} copies the data into a buffer on a page shared between the \gls{VM} and the \gls{HV}. 
Then, the \gls{VM} copies the data from the shared page to its destination in the \gls{VM}'s private memory.  
During this operation, the memory controller automatically encrypts the data when writing it to private memory. 

Consequently, data bounced between the \gls{VM}'s shared and private memory is accessible by the \gls{HV}. 
This leads to a window of opportunity for adversaries to inspect or modify the transferred information.  
Throughout this paper, we abuse the bounce buffering mechanism to inject arbitrary code into the \gls{VM}'s memory, without having to know the 
\gls{VM}'s encryption key. 
By combining such injections with \gls{SEV-ES}' lack of memory integrity protection, we defeat AMD's efforts against code execution attacks.

\section{Adversarial Capabilities}
\label{sec:threat}
To mount the \name attack, we assume that the attacker is in control of the \gls{HV}.
Whether the \gls{HV} has been compromised or intentionally set up for malicious purposes is out of scope.  
The goal is to inject and execute arbitrary code in \glspl{VM}, whose memory is encrypted by 
the latest AMD \gls{SEV-ES} technology.  
This architecture generally forbids executing any code from shared,
non-encrypted memory~\cite{AMD2020}, 
preventing the injection and execution of unencrypted payloads.
Further, the attacker is neither in the possession of the secret key nor has expert knowledge about the applied encryption algorithm. 
As such, the adversary cannot encrypt the payload before injecting it into the \gls{VM}.
The attacker is also not aware of any flaws in the utilized tweak function,
which prevents moving cipher-text blocks in memory~\cite{du2017secure,
li2019exploiting, wilke2020sevurity} (\autoref{sec:background:memenc}). 
In other words, contrary to \emph{SEVurity}~\cite{wilke2020sevurity}, the attacker is not concerned about the exact CPU version. 

Even though the attacker does not require the source code of the kernel running inside the
\gls{VM}, we do assume %
access to the \gls{VM}'s kernel, or the kernel's exported symbol information.  
Inside the target \gls{VM}, 
we expect a Linux kernel with \gls{KASLR}. 
Also, we expect the \gls{VM} to employ virtio for communicating with the system's devices. %
Both assumptions do not rely on any specific version of the Linux kernel or virtio. 
Further, disregarding the focus of
the prototype presented in this work, the attacker is not limited to exploiting
the network communication channel. 
In this context, the attacker disregards any in-\gls{VM} IP or port filtering mechanisms, and is also 
agnostic to any in-\gls{VM} services or open ports~\cite{hetzelt2017security, du2017secure, morbitzer2018severed, morbitzer2019extracting, li2019exploiting}.
The attacker can inject truly arbitrary payloads into the \gls{VM}'s kernel, and therefore does not have to consider any kernel-level defenses that target attacks from user space. 
Despite knowing the addresses of the kernel's symbols, the attacker will have to determine the randomized kernel base address, imposed by \gls{KASLR}~\cite{kaslr_edge:2013}. 
Finally, the attacker relies 
on AMD's event injection interface to execute virtual interrupts at known registered locations.

\section{The SEVerity Code Execution Attack}
\label{sec:attack}
\glsreset{SLAT}
\glsreset{GFN}

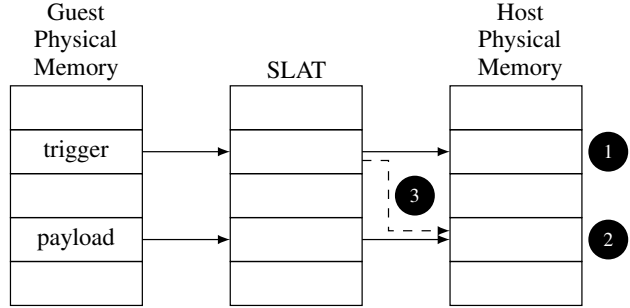
\begin{figure}[t]
    \centering
     \resizebox{1\columnwidth}{!}{%
       \begin{tikzpicture}[font=\footnotesize]
  \tikzset{>=latex}
  
  \draw[] (0,0) rectangle (1.5,0.5);
  \draw[] (0,0.5) rectangle (1.5,1) node[pos=.5] {payload};
  \draw[] (0,1) rectangle (1.5,1.5);
  \draw[] (0,1.5) rectangle (1.5,2.0) node[pos=.5] {trigger};
  \draw[] (0,2) rectangle (1.5,2.5); 
  
  \draw[] (2.5,0) rectangle (4,0.5);
  \draw[] (2.5,0.5) rectangle (4,1);
  \draw[] (2.5,1) rectangle (4,1.5);
  \draw[] (2.5,1.5) rectangle (4,2);
  \draw[] (2.5,2.0) rectangle (4,2.5);
  
  \draw[] (5,0) rectangle (6.5,0.5);
  \draw[] (5,0.5) rectangle (6.5,1);
  \draw[] (5,1) rectangle (6.5,1.5);
  \draw[] (5,1.5) rectangle (6.5,2);
  \draw[] (5,2) rectangle (6.5,2.5);
  
  \draw[->] (1.5,0.75) -- (2.5,0.75);
  \draw[->] (1.5,1.75) -- (2.5,1.75);
  \draw[->] (4,0.75) -- (5,0.75);
  \draw[->] (4,1.75) -- (5,1.75);
  \draw[->,dashed] (4,1.65) -- (4.3,1.65) -- (4.3,0.85) -- (5,0.85); 
  
  \node at (0.75,3.0) {\begin{tabular}{c} Guest\\Physical\\Memory\end{tabular}};
  \node at (3.25,2.65) {\begin{tabular}{c} SLAT \end{tabular}};
  \node at (5.8,3.0) {\begin{tabular}{c} Host\\Physical\\Memory\end{tabular}};
 
 \node[circle,fill=black,text=white,scale=0.8] at (6.8, 1.75) {1};
 \node[circle,fill=black,text=white,scale=0.8] at (6.8, 0.75) {2};
 \node[circle,fill=black,text=white,scale=0.8] at (4.6,1.25) {3};
   
\end{tikzpicture}
     }
     \caption{Overview of the \name attack. Having determined a trigger (\phase{1}), we send a payload to the \gls{VM} (\phase{2}).
                    After having determined the location of the payload, we modify the SLAT tables (\phase{3}) to trigger the execution of the payload. }
     \label{fig:attack}
\end{figure}

The \name attack challenges the security model of AMD's state-of-the-art \gls{VM} memory encryption feature \gls{SEV-ES}~\cite{sev-es}. 
For the attack, we expect the attacker to control the \gls{HV}, recreating AMD's original attacker model, against which \gls{SEV-ES} is intended to protect. %
This control allows the attacker to abuse architectural properties of \gls{SEV-ES} to mount \emph{arbitrary code execution attacks} against \glspl{VM} in different memory isolation domains, despite the added confidentiality guarantees.

\smallskip \noindent
Our attack comprises the following three steps (\autoref{fig:attack}), which we elaborate in detail in this section:
\begin{itemize}[nosep]
    \item[\phase{1}] Identify a code region inside the \gls{VM}, the execution of which we can trigger from the outside.
    \item[\phase{2}] Send a network packet with malicious payload to the \gls{VM} and identify its location inside the \gls{VM}'s
        memory.
    \item[\phase{3}] Modify the \gls{SLAT} tables to point the identified trigger to the injected payload and trigger its execution.
\end{itemize}

To identify our trigger (\phase{1}), we remove the \texttt{execute} flag from all of the \gls{VM}'s physical memory pages, the \glspl{GFN}, and inject a \gls{NMI} into the \gls{VM}. 
This allows us to determine the \glspl{GFN} that hold the \gls{NMI} handler and the
\texttt{do\_nmi()} function, the latter of which we intend to use as trigger.
To identify the location of network packets  within the \gls{VM}'s memory (\phase{2}), we track the write accesses of the \gls{VM} when copying the packet from the shared into private memory. 
Combined with detailed knowledge about virtio, this allows us to determine the exact location of every incoming network packet within the \gls{VM}'s memory. 
To conclude our attack, we inject a network packet containing our payload at a carefully calculated offset into the \gls{VM}. 
Afterwards, we abuse the lack of integrity protection and modify the \gls{SLAT} tables managing the mapping between \glspl{GFN} and \glspl{SFN} (\phase{3}). 
Specifically, we map the \gls{GFN} containing the trigger to the \gls{SFN} of the payload. 
Finally, by injecting an \gls{NMI} into the \gls{VM}, we trigger the execution of our payload.

Note that we do not limit \name to virtio or Linux. %
Even though our work uses Linux and virtio as concrete examples to mount the \name attack,
the pillars upon which we base the attack are completely \gls{OS} and subsystem (i.e.,
driver) independent.
Thus, targeting another \gls{OS} or subsystem would not change the essence of the three steps \mbox{(\phase{1} -- \phase{3})}, which
form the \name attack (\autoref{sec:discussion}). 

\subsection{Identifying the Trigger Point}
\label{sec:attack:code}

The virtualization extensions of AMD allow the system to pass interrupts directly to the \gls{VM}.
Alternatively, the \gls{HV} can 
intercept an interrupt and subsequently inject a \emph{virtual interrupt} into the \gls{VM} through the \emph{event injection} interface~\cite{AMD2020}. 
Just as real interrupts, virtual interrupts cause \glspl{VM} to pause the currently executing code and handle the incoming event urgently in the associated interrupt handler.  
Consequently, virtual interrupts present themselves as good candidates that we can abuse for triggering code inside a \gls{VM}.
Yet, most interrupts are vital to the \gls{VM}'s execution and temporary replacing the interrupt's code with our payload would crash the \gls{VM}.
Therefore, we 
resort to one specific interrupt type, the \gls{NMI}, that is rarely generated. %

The CPU uses the
Interrupt Descriptor Table, 
a hardware-defined data structure that registers all interrupt vectors, to locate the \gls{NMI} handler.
Once the \gls{HV} injects a virtual \gls{NMI} into a \gls{VM}, the hardware immediately interrupts the \gls{VM}'s execution and invokes the \gls{NMI} handler. 
In the context of \name, we can abuse this 
behavior
to initiate the execution of arbitrary code residing at the location of the registered \gls{NMI} handler.
Specifically, if we remap the \gls{GFN} of the \gls{NMI} handler to the \gls{SFN} containing our payload (\autoref{fig:attack}) and issue a virtual \gls{NMI}, we will be able to execute our payload.

Consequently, in order to perform the remapping, we have to identify the \gls{GFN} of the \gls{NMI} handler (\phase{1}).
For this, we assume that the \gls{HV} has access to the exported symbol information or to the unencrypted image of the \gls{VM}'s kernel.
This is a valid assumption, as \gls{VM} images are often provided by the cloud provider and the unencrypted kernel image usually resides in \texttt{/boot}. 
By analyzing the kernel image, we can determine the offset of the \gls{NMI} handler. 
We discuss alternatives for determining the offset in Section \autoref{sec:discussion}.
However, Linux %
applies \gls{KASLR} to randomize the
offset of the kernel image 
in the \gls{VM}'s virtual and physical address space~\cite{kaslr_edge:2013, kernel2020self}.
Fortunately, we are able to 
use 
different approaches that allow us to bypass
\gls{KASLR} and hence determine the \gls{GFN} of the \gls{NMI} handler. For
instance, we can employ any of the following methods:

\smallskip \noindent
\textbf{Manipulating \gls{KASLR}:}
On modern AMD platforms, the \gls{TSC} counts the number of processor-clock cycles since the last reset~\cite{AMD2020}, which 
can be read via the \texttt{RDTSC} instruction. 
The Linux kernel uses this instruction
to create entropy at boot time, which is among others required to initialize \gls{KASLR}. 
Therefore, a malicious \gls{HV} manipulating the results of the \texttt{RDTSC} instruction at boot time will be able to pin the \gls{KASLR} offset~\cite{radev2020abusing}, allowing us
to guess the \gls{KASLR} offset used by the \gls{VM}. 
However, this method requires a malicious \gls{HV} to actively interfere with the \gls{VM} at boot time. 
Additionally, the \gls{VM} would be able to detect the static \gls{KASLR} offset. 
To circumvent these limitations, we can also 
use 
techniques which allow us to infer the \gls{KASLR} offset at runtime. 

\noindent%
\textbf{Fingerprinting nested page faults:} %
Wilke et al.~\cite{wilke2020sevurity} observed that
nested page faults occur \emph{deterministically} during the 
boot process.
Specifically, in the initial stage,
before the \gls{VM}'s \gls{OS} activates
\gls{KASLR}, the \gls{VM} exhibits the same \gls{GFN} access pattern across reboots.
Only once the boot process starts to decompress
the kernel image to a random offset in the \gls{VM}'s physical memory, the
sequence will start to differ. We have determined that the first \gls{GFN}
which differs from this sequence holds the beginning of the kernel image in the
\gls{VM}'s physical memory.  
As the beginning of the kernel image is always aligned on a $2$MB boundary~\cite{schwarz2019store}, we can safely assume that the start of the image will always remain at page offset zero. 
This method allows the \gls{HV} to passively analyze the \gls{GFN} access pattern of
the \gls{VM}'s early boot stage. %
Yet, since the attacker does not necessarily have the chance to observe the bootstrapping phase of the \gls{VM}, we have developed the following alternative method.

\noindent%
\textbf{Probing the \gls{NMI} handler:} 
We can leverage the fact that we control the \gls{SLAT} tables to identify the
\gls{GFN} of the \gls{NMI} handler. Specifically, by revoking the
\texttt{execute} 
permission from all \glspl{GFN} in the
\gls{SLAT} tables, we can track all execution attempts of the \gls{VM}.
This configuration of the \gls{SLAT} tables causes the \gls{VM} to trap into the \gls{HV} on every instruction fetch.
For every trap, we can determine the \gls{GFN} which caused the trap.
That is, if we apply this configuration and inject a virtual \gls{NMI} into the
\gls{VM}, 
the \gls{VM} will attempt to execute the registered \gls{NMI} handler and immediately trap into the \gls{HV}. 
Using this method, we are able to locate the \gls{GFN} of the \gls{NMI} handler.
In order to continue the \gls{VM}'s execution, we restore the permissions of the \gls{GFN} containing the \gls{NMI} handler and resume the \gls{VM}.
Shortly after resuming the \gls{VM}, it will trap again upon calling the
\texttt{do\_nmi()} function 
responsible for dispatching \gls{NMI}'s.
Since \texttt{do\_nmi()} resides in a different section
(\texttt{.text}) than the \gls{NMI} handler itself  (\texttt{.entry.text}), we
can be sure that the call to \texttt{do\_nmi()} will trap into the
\gls{HV}.
In other words, both functions reside on different \glspl{GFN} and thus
guarantee two consecutive traps.

\smallskip

Once we have identified the \gls{GFN} holding the \texttt{do\_nmi()} function, all that is left to determine is the function's exact offset within the respective \gls{GFN}.
This knowledge allows us to inject our payload at the same offset, ensuring it will be immediately executed after we modify the \gls{SLAT} tables and trigger its execution.  
To determine the offset of the \texttt{do\_nmi()} function, 
we analyze the exported symbol information of the \gls{VM}'s kernel image. 

\subsection{Identifying the Payload Destination}
\label{sec:attack:data}
The architectural constraints of \gls{SEV-ES} prohibit devices to apply \gls{DMA} into encrypted \gls{VM} memory (\autoref{sec:background:dma}). 
This limitation applies to physical as well as emulated devices, forcing \glspl{VM} and devices to exchange data through shared pages.
Even though this security-driven design decision aims to protect the system against \gls{DMA} attacks~\cite{becher2005firewire}, it also establishes a vantage point:
the unprotected \gls{DMA} channel facilitates attackers to 
$(i)$~extract or inject arbitrary data from or to the \gls{VM}, or
$(ii)$~inject arbitrary \emph{code} into the \gls{VM}'s memory without knowing the secret encryption key (\autoref{sec:background:memenc}). 
While previous attacks have targeted the former~\cite{li2019exploiting}, the latter attack vector 
is the main focus of \name.  
Specifically, we abuse the exposed network communication channel to inject arbitrary code into the encrypted memory.
Yet, before injecting the payload, we need to identify where in memory the \gls{VM} stores incoming network packets (\phase{2}). 

\subsubsection{Virtio tracking}

The goal of virtio tracking is to determine the \glspl{GFN} in which the \gls{VM} maintains incoming network packets. 
For this, we record the sequence of \glspl{GFN} the \gls{VM} is writing to when bouncing an incoming network packet from shared into private memory. 
To determine which of the \glspl{GFN} contains which network packet, we need to understand how \gls{SEV-ES} protected \glspl{VM} use virtio~\cite{oasis2019virtio} to exchange data between the virtual device in the \gls{HV} and the driver in the \gls{VM}.
Virtio uses \texttt{virtqueues} as its main communication structures.
Each virtual network device uses at least two \texttt{virtqueues}, one for receiving data (\texttt{rx virtqueue}), and one for sending data (\texttt{tx virtqueue}).
The \texttt{virtqueues} can be \texttt{packed} or \texttt{split}. 
For simplicity reasons, we will focus on \texttt{split virtqueues}, which were used by all \glspl{VM} we analyzed.
Each \texttt{split virtqueue} comprises three components: $(i)$~the \emph{descriptor table}, $(ii)$~the \emph{available ring}, and $(iii)$~the \emph{used ring}. 
Each column of the descriptor table contains, among others, the address and length of a buffer the device and the driver can use for exchanging data. 
To refer to a buffer,
the driver and the device use the buffer's position in the descriptor table as an index. 
The \gls{VM}'s driver writes indices into the available ring, offering buffers to the virtual device in the \gls{HV}. 
Once the virtual device has used one or more of the available buffers, it adds the buffer's index to the used ring. 

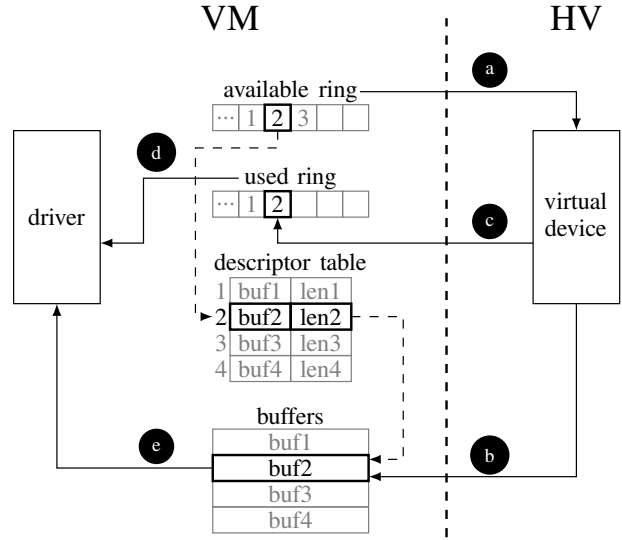
\begin{figure}[t]
    \centering
    \resizebox{1\columnwidth}{!}{%
       \begin{tikzpicture}[font=\footnotesize]
  \tikzset{>=latex}
  
  \draw[] (0,0) rectangle (1,2);
  \draw[] (6,0) rectangle (7,2);
  \node at (0.5,1) {driver};
  \node at (6.5,1) {\begin{tabular}{c} virtual\\device\end{tabular}};
  
  \node at (2.5,3.30) {\large VM};
  \node at (6.5,3.30) {\large HV};
  
  \draw[dashed, thick] (5,-2.7) -- (5,3.3);
  
  \node at (3.2,2.45) {available ring};
  
  \draw[gray] (2.3,2) rectangle (2.6,2.3) ;
  \draw[gray] (2.6,2) rectangle (2.9,2.3);
  \draw[gray] (3.2,2) rectangle (3.5,2.3);
  \draw[gray] (3.5,2) rectangle (3.8,2.3);
  \draw[gray] (3.8,2) rectangle (4.1,2.3);
  \draw[thick] (2.9,2) rectangle (3.2,2.3);
  
  \node[gray] at (2.45,2.15) {...};
  \node[gray] at (2.75,2.15) {1};
  \node[] at (3.05,2.15) {2};
  \node[gray] at (3.35,2.15) {3};

  \node at (3.2,1.45) {used ring};

  \draw[gray] (2.3,1) rectangle (2.6,1.3) ;
  \draw[gray] (2.6,1) rectangle (2.9,1.3);
  \draw[gray] (3.2,1) rectangle (3.5,1.3);
  \draw[gray] (3.5,1) rectangle (3.8,1.3);
  \draw[gray] (3.8,1) rectangle (4.1,1.3);
  \draw[thick] (2.9,1) rectangle (3.2,1.3);
  
  \node[gray] at (2.45,1.15) {...};
  \node[gray] at (2.75,1.15) {1};
  \node[] at (3.05,1.15) {2};

  \node[] at (3.2,0.45) {descriptor table};
  
  \draw[gray] (2.5,0) rectangle (3.2,0.3);
  \draw[gray] (3.2,0) rectangle (3.9,0.3);
  \draw[gray] (2.5,-0.6) rectangle (3.2,-0.3);
  \draw[gray] (3.2,-0.6) rectangle (3.9,-0.3);
  \draw[gray] (2.5,-0.9) rectangle (3.2,-0.6);
  \draw[gray] (3.2,-0.9) rectangle (3.9,-0.6);
  \draw[thick] (2.5,-0.3) rectangle (3.2,0);
  \draw[thick] (3.2,-0.3) rectangle (3.9,0);
  
  \node[gray] at (2.4,0.15) {1};
  \node[gray] at (2.85,0.15) {buf1};
  \node[gray] at (3.55,0.15) {len1};
  \node[gray] at (2.4,-0.45) {3};
  \node[gray] at (2.85,-0.45) {buf3};
  \node[gray] at (3.55,-0.45) {len3};
  \node[gray] at (2.4,-0.75) {4};
  \node[gray] at (2.85,-0.75) {buf4};
  \node[gray] at (3.55,-0.75) {len4};
  \node[] at (2.4,-0.15) {2};
  \node[] at (2.85,-0.15) {buf2};
  \node[] at (3.55,-0.15) {len2};
  
  \node[] at (3.2,-1.3) {buffers};

  \draw[gray] (2.3,-1.75) rectangle (4.1,-1.45);  
  \draw[gray] (2.3,-2.35) rectangle (4.1,-2.05);  
  \draw[gray] (2.3,-2.65) rectangle (4.1,-2.35);  
  \draw[thick] (2.3,-2.05) rectangle (4.1,-1.75);  
  
  \node[gray] at (3.2,-1.6) {buf1};
  \node[] at (3.2,-1.9) {buf2};
  \node[gray] at (3.2,-2.2) {buf3};
  \node[gray] at (3.2,-2.5) {buf4};

  \draw[->] (4.0,2.45) -- (6.5,2.45) -- (6.5,2);
  \node[circle,fill=black,text=white,scale=0.8] at (5.5, 2.70) {a};
  
  \draw[->] (6.5,0) -- (6.5,-2.0) -- (4.1,-2.0);
  \node[circle,fill=black,text=white,scale=0.8] at (5.5, -1.75) {b};
  
  \draw[->] (6,0.7) -- (3.05,0.7) -- (3.05,1);
  \node[circle,fill=black,text=white,scale=0.8] at (5.5, 0.95) {c};
  
  \draw[->] (2.6,1.45) -- (1.5,1.45) -- (1.5,0.7) -- (1,0.7);
  \node[circle,fill=black,text=white,scale=0.8] at (1.65, 1.75) {d};
  
  \draw[->] (2.3,-1.9) -- (0.5,-1.9) -- (0.5,0);
  \node[circle,fill=black,text=white,scale=0.8] at (1.65, -1.65) {e};
  
  \draw[dashed,->] (3.05, 2) -- (3.05, 1.8) -- (2.1,1.8) -- (2.1,-0.15) -- (2.3,-0.15);
  \draw[dashed,->] (3.9,-0.15) -- (4.5,-0.15) -- (4.5,-1.8) -- (4.1,-1.8);
\end{tikzpicture}
    }
    \caption{Abstract virtio network architecture. The virtual device in the \gls{HV} leverages the \emph{available ring buffer} and the \emph{descriptor table} to locate %
the buffer for %
incoming packets. In the \gls{VM}, the driver uses the \emph{used ring buffer} and the \emph{descriptor table} to determine the buffers holding the packets.}
    \label{fig:virtio}
\end{figure}

\autoref{fig:virtio} depicts the process of an incoming networking packet using the \texttt{rx virtqueue}.
To enable the \gls{VM} to receive network packets, the driver allocates buffers, adds them to the descriptor table and their respective index into the available ring. 
When a network packet comes in, the virtual device reads the available ring to find the next free buffer (\circled{a}) and writes the packet into 
it
(\circled{b}).
Then, the device adds the buffer's index to the used ring (\circled{c}). 
By reading the used ring (\circled{d}), the driver in the \gls{VM} 
determines from which buffer to consume the network packet (\circled{e}).

The described method requires the virtual device to access different parts of the \texttt{virtqueue}.
As it is managed by the driver in the \gls{VM}, the buffers are located within the \gls{VM}'s memory region. 
However, when using \gls{SEV-ES}, the virtual device in the \gls{HV} is not able to read or write meaningful data
due to the encrypted memory. 
\gls{SEV-ES} works around this limitation by using bounce buffers (\autoref{sec:background:dma}). 

Using bounce buffers, the \gls{VM} allocates the \texttt{virtqueue} structure in a memory region shared between the \gls{VM} and the \gls{HV}. 
After the device puts a packet into a shared buffer, the \gls{VM} bounces the 
buffer's content
from the shared, unencrypted memory into the private, encrypted memory. 

We track page accesses
to determine to which location in %
private memory the \gls{VM} bounces incoming network packets. 
Specifically, after the device in the \gls{HV} places a network packet into a buffer in shared memory, we modify the \gls{SLAT} tables by removing the \texttt{present} flag from the \gls{GFN} holding the buffer. 
Hence, the \gls{HV} gets notified as soon as the \gls{VM} tries to access the buffer. %
Having received this notification, we conclude that the \gls{VM} is about to read the buffer. 
Next, we enable \emph{write tracking} on all other \glspl{GFN} by revoking their write access permission in the \gls{SLAT} tables. 
This allows us to collect the sequence of \glspl{GFN} the \gls{VM} is writing to until 
it has finished bouncing the packet into the private memory. 

We abuse virtio's notification system to determine when the \gls{VM} has finished bouncing the packet.
The \gls{HV} raises an interrupt to notify the driver that it can consume a buffer. 
Yet, if the driver is busy,
the virtio specification allows the driver in the \gls{VM} to let the \gls{HV} know that it does not want to receive interrupts. 
In older virtio implementations, the driver informs the device by setting the \texttt{VRING\_AVAIL\_F\_NO\_INTERRUPT} flag;
the driver will unset this flag once it has finished processing the data. 
In recent virtio implementations, 
once the driver has finished consuming a buffer, it updates the \texttt{used\_event} to the index of the consumed buffer. 
We monitor the \texttt{VRING\_AVAIL\_F\_NO\_INTERRUPT} flag 
and
the \texttt{used\_event} entry in order to comply with all virtio versions. 
This allows us to determine 
when the driver has completed processing the packet. 

\subsubsection{Packet buffer tracking}

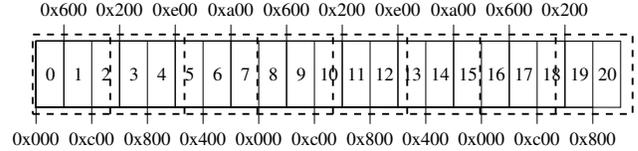
\begin{figure}[t]
    \centering
    \resizebox{1\columnwidth}{!}{%
       \begin{tikzpicture}[font=\footnotesize]
  \tikzset{>=latex}
  
  \draw[] (0,0) rectangle (.42,1);
  \draw[] (0,0) rectangle (.84,1);
  \draw[] (0,0) rectangle (1.26,1);
  \draw[] (0,0) rectangle (1.68,1);
  \draw[] (0,0) rectangle (2.10,1);
  \draw[] (0,0) rectangle (2.52,1);
  \draw[] (0,0) rectangle (2.94,1);
  \draw[] (0,0) rectangle (3.36,1);
  \draw[] (0,0) rectangle (3.78,1);
  \draw[] (0,0) rectangle (4.20,1);
  \draw[] (0,0) rectangle (4.62,1);
  \draw[] (0,0) rectangle (5.04,1);
  \draw[] (0,0) rectangle (5.46,1);
  \draw[] (0,0) rectangle (5.88,1);
  \draw[] (0,0) rectangle (6.30,1);
  \draw[] (0,0) rectangle (6.72,1);
  \draw[] (0,0) rectangle (7.14,1);
  \draw[] (0,0) rectangle (7.56,1);
  \draw[] (0,0) rectangle (7.98,1);
  \draw[] (0,0) rectangle (8.40,1);
  \draw[] (0,0) rectangle (8.82,1);
  
  \node at (.22,.5) {0};
  \node at (.64,.5) {1};
  \node at (1.06,.5) {2};
  \node at (1.48,.5) {3};
  \node at (1.90,.5) {4};
  \node at (2.32,.5) {5};
  \node at (2.74,.5) {6};
  \node at (3.16,.5) {7};
  \node at (3.58,.5) {8};
  \node at (4.00,.5) {9};
  \node at (4.42,.5) {10};
  \node at (4.84,.5) {11};
  \node at (5.26,.5) {12};
  \node at (5.68,.5) {13};
  \node at (6.10,.5) {14};
  \node at (6.52,.5) {15};
  \node at (6.94,.5) {16};
  \node at (7.36,.5) {17};
  \node at (7.78,.5) {18};
  \node at (8.20,.5) {19};
  \node at (8.62,.5) {20};
  
  \draw[] (0,0) -- (0,-.25) node[below] {0x000};
  \draw[] (.84,0) -- (.84,-.25) node[below] {0xc00};
  \draw[] (1.68,0) -- (1.68,-.25) node[below] {0x800};
  \draw[] (2.52,0) -- (2.52,-.25) node[below] {0x400};
  \draw[] (3.36,0) -- (3.36,-.25) node[below] {0x000};
  \draw[] (4.20,0) -- (4.20,-.25) node[below] {0xc00};
  \draw[] (5.04,0) -- (5.04,-.25) node[below] {0x800};
  \draw[] (5.88,0) -- (5.88,-.25) node[below] {0x400};
  \draw[] (6.72,0) -- (6.72,-.25) node[below] {0x000};
  \draw[] (7.56,0) -- (7.56,-.25) node[below] {0xc00};
  \draw[] (8.40,0) -- (8.40,-.25) node[below] {0x800};
  
  \draw[] (.42,0) -- (.42,1.25) node[above] {0x600};
  \draw[] (1.26,0) -- (1.26,1.25) node[above] {0x200};
  \draw[] (2.10,0) -- (2.10,1.25) node[above] {0xe00};
  \draw[] (2.94,0) -- (2.94,1.25) node[above] {0xa00};
  \draw[] (3.78,0) -- (3.78,1.25) node[above] {0x600};
  \draw[] (4.62,0) -- (4.62,1.25) node[above] {0x200};
  \draw[] (5.46,0) -- (5.46,1.25) node[above] {0xe00};
  \draw[] (6.30,0) -- (6.30,1.25) node[above] {0xa00};
  \draw[] (7.14,0) -- (7.14,1.25) node[above] {0x600};
  \draw[] (7.98,0) -- (7.98,1.25) node[above] {0x200};
  
  \draw[thick,dashed] (-0.05,-.1) rectangle (9,1.1);
  \draw[thick,dashed] (1.12,-.1) -- (1.12,1.1);
  \draw[thick,dashed] (2.24,-.1) -- (2.24,1.1);
  \draw[thick,dashed] (3.34,-.1) -- (3.34,1.1);
  \draw[thick,dashed] (4.48,-.1) -- (4.48,1.1);
  \draw[thick,dashed] (5.60,-.1) -- (5.60,1.1);
  \draw[thick,dashed] (6.70,-.1) -- (6.70,1.1);
  \draw[thick,dashed] (7.84,-.1) -- (7.84,1.1);

\end{tikzpicture}
    }
  \caption{
      The $32$KiB sized \texttt{packet buffer} contains $21$ fragments of \texttt{0x600} bytes. 
      The size of $32$KiB requires $8$ \glspl{GFN}, indicated by the dashed lines. 
  }
  \label{fig:packetbuffer}
\end{figure}

Having determined the write sequence when bouncing a network packet into the \gls{VM}, we identify which \gls{GFN} in the sequence contains the packet, and at which offset the packet is located. 
To determine the exact position of a packet, we have to understand how virtio manages its buffers. 
The buffers are allocated in $32$KiB chunks of (physically) contiguous memory called a \texttt{packet buffer}.
Each \texttt{packet buffer} is divided into multiple fragments of the same size. 
\autoref{fig:packetbuffer} shows the layout of a \texttt{packet buffer} holding $21$ fragments of \texttt{0x600} bytes each. 
The fragments are aligned and padded to \texttt{0x600} bytes, with the first fragment starting at offset \texttt{0x0}. 
The dashed lines indicate the eight \glspl{GFN} holding the buffer. 
 
Note that within one \texttt{packet buffer}, all fragments have the same size. 
Yet, the 
fragment size 
 can be dynamically adjusted to consider the size of recent network packets~\cite{dalton2013mergeable}. 
Therefore, the number of fragments per \texttt{packet buffer} can vary. 
On our \glspl{VM}, we have observed a fragment size of \texttt{0x600} bytes. 
This size can change when the \gls{VM} starts to receive larger frames.
We can detect changes to the size of the fragments in the \texttt{packet buffer}
by monitoring the length of the buffers in the descriptor table. %

This knowledge of the \texttt{packet buffer}'s structure allows us to infer which \glspl{GFN} contain which fragments, and which fragment starts at which offset.
First,
we know that the \texttt{packet buffer} is $32$KiB aligned in the \gls{VM}'s memory. 
Therefore, we search the write sequences performed by the \gls{VM} when bouncing a packet into private memory for a 
$32$KiB aligned \gls{GFN}.
This \gls{GFN} represents a candidate for the first \gls{GFN} of a newly allocated \texttt{packet buffer}. 
Knowing that the first \gls{GFN} contains the start of three fragments (\autoref{fig:packetbuffer}), we confirm the candidate by analyzing if the \gls{VM} also writes to the \gls{GFN} when bouncing the 
next
two packets into private memory. 

We can repeat this process for the subsequent \glspl{GFN} in the \texttt{packet buffer} to minimize false positives. 
For this, we 
again use
our detailed knowledge of the \texttt{packet buffer}. 
For example, the second \gls{GFN} of the \texttt{packet buffer} contains the start of fragments $3$, $4$ and $5$. 
Therefore, assuming we have correctly identified the first \gls{GFN} of a packet buffer, the second, subsequent \gls{GFN} should be accessed in the fourth, fifth, and sixth bouncing sequence.
The third \gls{GFN} containing the start of fragments $6$ and $7$ should be accessed in the sixth and seventh bouncing sequence, and so on. 
  
Note that depending on the size of the incoming network packet, the \gls{VM} may also access the second \gls{GFN} in the third sequence.  
Considering the minimum size of an Ethernet frame (\texttt{0x40} bytes)~\cite{ieee2018ethernet}, the packet stored in the third fragment will not cause a write access to the second \gls{GFN}. 
Yet, a packet bigger than \texttt{0x400} bytes will cause the \gls{VM} to also write data to the second \gls{GFN}; the fragment offset (\texttt{0xc00}) plus size (\texttt{0x400}) cross the first \gls{GFN}'s boundary. 

By applying these conditions and adapting them 
to reduce the number of false positives, we are able to successfully identify the \texttt{packet buffer} within the \gls{VM}'s memory.  
Our tests show that requiring the first ten write sequences of the bounce buffer to fulfill our conditions eliminates the chance for false positives on all tested \glspl{VM}. 
Having determined the position of the \texttt{packet buffer}, we 
can also 
determine the offset of each fragment by knowing its size. 
This allows us to predict where the \gls{VM} will store prospective incoming network packets.

Note, the \gls{VM} copies incoming network packets into the \texttt{packet buffer} before it applies any processing or filtering.
Thus, the \gls{VM} stores packets sent to closed or even firewalled ports in the \texttt{packet buffer}. 
This allows us to send our payload to closed or filtered ports and to eliminate the need for any active network service running in the \gls{VM}.

\subsection{Injecting the Payload}
\label{sec:attack:payload}

Having determined the trigger point (\autoref{sec:attack:code}) and the location to which 
to inject our payload (\autoref{sec:attack:data}), we finalize the setup and conclude the \name attack (\phase{3}).
The final step of the attack comprises two stages. 
First, we inject a payload into a suitable fragment in the \texttt{packet buffer}.
Second, we manipulate the \gls{SLAT} tables to remap the trigger such that its \gls{GFN} translates to the \gls{SFN} of the injected payload.  
Finally, %
we inject an \gls{NMI} to execute the payload.

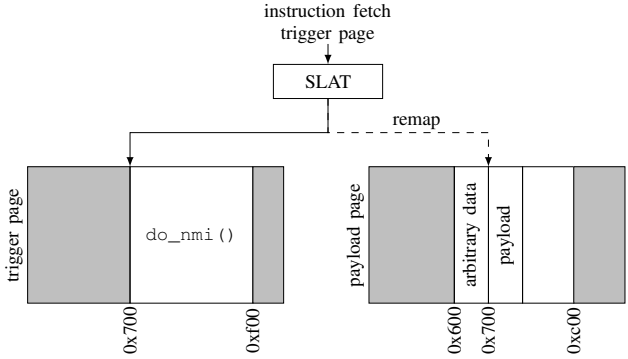
\begin{figure}[t]
    \centering
    \resizebox{1\columnwidth}{!}{%
       \begin{tikzpicture}[font=\footnotesize]
  \tikzset{>=latex}
  
  \node at (4.4,4.1) {\begin{tabular}{c}instruction fetch\\trigger page\end{tabular}};
  \draw[] (3.6,3) rectangle (5.2,3.5);

  \draw[fill=lightgray] (0,0) rectangle (1.5,2);
  \draw[] (1.5,0) rectangle (3.3,2);
  \draw[fill=lightgray] (3.3,0) rectangle (3.75,2);
  
  \draw[fill=lightgray] (5,0) rectangle (6.25,2);
  \draw[] (6.25,0) rectangle (6.75,2);
  \draw[] (6.75,0) rectangle (7.25,2);
  \draw[] (7.25,0) rectangle (8,2);
  \draw[fill=lightgray] (8,0) rectangle (8.75,2);
  
  \draw[->] (4.4,3.8) -- (4.4,3.5);
  \draw[->] (4.4,3) -- (4.4,2.5) -- (1.5, 2.5) -- (1.5,2);
  \draw[->, dashed] (4.4,3) -- (4.4,2.5) -- (6.75,2.5) -- (6.75,2);
  
  \node at (4.4,3.25) {SLAT};
  \node at (5.7,2.65) {remap};
  \node at (2.4,1) {\texttt{do\_nmi()}};
  \node[rotate=90] at (1.5,-0.4) {0x700};
  \node[rotate=90] at (3.3,-0.4) {0xf00};
  \node[rotate=90] at (6.25,-0.4) {0x600};
  \node[rotate=90] at (6.75,-0.4) {0x700};
  \node[rotate=90] at (8,-0.4) {0xc00};
  \node[rotate=90] at (-0.2,1) {trigger page};
  \node[rotate=90] at (4.8,1) {payload page};
  \node[rotate=90] at (6.5,1) {arbitrary data};
  \node[rotate=90] at (7,1) {payload};
\end{tikzpicture}
    }
  \caption{
      The system maps the \gls{GFN} of the \emph{trigger page} to the
      original \gls{SFN} holding \texttt{do\_nmi()}.
      \name leverages the \gls{SLAT} to remap the \gls{GFN} of the trigger page
      to another \gls{SFN} holding the injected payload. Thus, by injecting
      an \gls{NMI}, we 
      redirect the control-flow to our payload.
  }
  \label{fig:offsets}
\end{figure}

When injecting the payload into a suitable fragment, we have to consider that
the trigger and the fragment can reside on different page offsets. 
As such, we must place the payload into a fragment whose location in memory spans a range 
comprising
the needed page offset. 
In
\autoref{fig:offsets}, the trigger is located at the page offset \texttt{0x700}.
Thus, we 
inject the payload to a fragment 
starting 
at offset \texttt{0x600}. 
To position the payload at precisely the same offset as the trigger, we fill the
first \texttt{0x100} bytes of the network packet %
with arbitrary data. 
Since the fragment also holds protocol headers of the packet, 
the arbitrary data only spans the size of \texttt{0x100} bytes minus the size of the protocol headers (e.g., Ethernet frame, IP and TCP or UDP header). 
After a total of exactly \texttt{0x100} bytes, we place our payload to ensure that the payload and the trigger are located at the same page offset.

For the next step, we make use of the fact that while \gls{SEV-ES}' memory encryption protects the guest page tables within the \gls{VM}, the \gls{HV} remains in charge of the \gls{SLAT} tables. 
The \gls{SLAT} tables map the \gls{VM}'s physical memory pages, the \glspl{GFN}, to their actual location in the \gls{HV}'s physical memory, the \glspl{SFN}. 
Therefore, once we prepared and injected the payload, we adjust the \gls{SLAT} table entry such that the \gls{GFN} that initially translated to the \gls{SFN} holding the trigger (\autoref{sec:attack:code}) translates to the \gls{SFN} holding the injected payload (\autoref{sec:attack:data}).  
That is, once the \gls{NMI} handler attempts to call the \texttt{do\_nmi()} function, it will instead execute our payload.

\section{Evaluation}
\label{sec:evaluation}

To show that \name is effective and OS-agnostic, we created a \gls{PoC} and evaluated its performance on different \glspl{VM}.

\subsection{System Setup}
\label{sec:evaluation:setup}
Our setup comprises a Debian 11 host running a custom-built Linux kernel \texttt{v5.6}
including the unofficial \gls{SEV-ES} patches~\cite{sev-es}.
The host equips an 8-core EPYC 3251 CPU with 64\,GB of
RAM and uses QEMU \texttt{v4.2.50} and KVM to virtualize the \glspl{VM} to be
attacked. Further, we have extended the host's kernel with the
SEVered~\cite{severedframework} patches and the \name implementation to provide all
the functionality necessary for the attack,
such as functionality required to identify the \texttt{packet buffer}.  

We have evaluated \name by targeting \gls{SEV-ES} protected \glspl{VM} running the following five different
Linux distributions officially supported by AMD~\cite{AMDSEV}:
\gls{SLES} $15$; \gls{RHEL} $8$; Fedora $29$; Ubuntu $18.04$; and
openSUSE-Tumbleweed.
To enable \gls{SEV-ES}, all \glspl{VM} were running a custom built kernel in
version \texttt{v5.7.0} with patches for \gls{SEV-ES} support~\cite{sev-es}.

\subsection{Precision Evaluation}
\label{sec:evaluation:perf}

To evaluate 
\name's precision, 
we ran 
it
 $1000$ times against all five \glspl{VM}.
Before every round of the attack, we sent a random number between $0$ and $64$ packets with random data to avoid any bias from previous rounds. 
Afterwards, in every round of the attack, we identified the \gls{GFN} of \texttt{do\_nmi()} by injecting an \gls{NMI} and performing execute tracking (\autoref{sec:attack:code}). 
To identify the \texttt{packet buffer} in the \gls{VM}'s memory, we sent dummy packets and analyzed the page fault sequence for bouncing the packets into the \gls{VM}'s private memory (\autoref{sec:attack:data}).
Once a series of ten page fault sequences fulfilled our conditions, we deemed the \texttt{packet buffer} identified (\autoref{sec:attack:data}). 
While a smaller number may have increased the speed of the evaluation, the higher number allowed us to eliminate false positives. 
Having identified the \texttt{packet buffer}, we sent additional dummy packets 
if
the next free fragment was not located on a suitable offset to inject our payload (\autoref{sec:attack:payload}). 
The payload itself executed a hypercall with exit reason \texttt{0xff}, and afterwards issued a \texttt{RET} in order to return to the \gls{NMI} handler. 
Having performed 
our
preparations, we remapped the \gls{GFN} of \texttt{do\_nmi()} to the \gls{SFN} of the payload and triggered its execution. 
By monitoring for hypercalls with exit reason \texttt{0xff}, we determined successful execution of the payload.  
To conclude the round, we connected to the \gls{VM} via SSH to ensure that it continued to behave as expected.  

To perform the evaluation itself, we filtered any external traffic to the \gls{VM} to maintain full control over incoming network packets. 
In a real-world setting, incoming packets would have allowed us to determine
the exact destination of the payload in a stealthy way, without having to inject
any artificial packets. Yet, to avoid having to rely on fluctuating numbers of
incoming network packets, which would have influenced the evaluation results, we
have opted for completely controlling the network traffic.

\begin{table}[t]
    \centering
    \caption{%
        Evaluation details of $1000$ runs of \name against different \glspl{VM} protected with \gls{SEV-ES}. The table further summarizes an
        averaged number of the tracked write accesses to
        memory when bouncing network packets; the number of the sent packets;
        and the duration of the attack. %
    }
    \resizebox{\linewidth}{!}{
        \begin{tabular}{l|c|c|c|c}
            \toprule
            \textbf{VM image}			& \textbf{\#~Write Accesses}	& \textbf{\#~Sent Packets} 			& \textbf{Duration} 	& \textbf{Success} \\ 
            \midrule
            SLES                 	& 38.52 			& 19.97			& 2.57 \emph{sec}		& 100\,\% \\
            RHEL                	& 66.78			& 19.50			& 2.75 \emph{sec}		& 100\,\% \\
            Fedora  			& 68.74			& 20.37			& 3.04 \emph{sec}		& 100\,\% \\
            Ubuntu             	& 43.10			& 19.89			& 2.65 \emph{sec}		& 100\,\% \\
            openSUSE         	& 40.57			& 19.56			& 2.57 \emph{sec}		& 100\,\% \\
            \bottomrule
        \end{tabular}
    }
    \label{tab:results}
\end{table}

For each \gls{VM}, we conducted our attack $1000$ times, achieving a total success rate of 100\,\%. 
\autoref{tab:results} shows the detailed results of our evaluation. 
In the second column, we show the average number of write accesses recorded when monitoring the \glspl{VM} bouncing a packet from shared to private memory. 
Their values range from $38.52$ for openSUSE to $68.74$ for Fedora. %

In the third column, we show the average number of packets we had to send before we were able to identify the \texttt{packet buffer} in the \gls{VM}'s private memory. 
To ensure 
to
always correctly identify the \texttt{packet buffer}, we required $10$ consecutive write sequences to match our conditions. 
In our test environments, all \glspl{VM} filled the \texttt{packet buffer} with $21$ fragments of \texttt{0x600} bytes.
Therefore, on average, we had to send $10$ packets before a new \texttt{packet buffer} was allocated.
The results in the third column match the expected value of these $10$ packets plus $10$ packets to ensure the \texttt{packet buffer} identification was correct. %

The fourth column shows the average time 
to identify the \texttt{packet buffer} (\phase{2}), inject our payload, remap and trigger its execution (\phase{3}). 
The values are almost all below three seconds, ranging from $2.57s$ on \gls{SLES} and openSUSE to $3.04s$ on Fedora. 
From these results, we conclude that while filtering external traffic during the attack may be noticeable,
the short duration will not attract specific attention. 
Additionally, extending our implementation to buffer incoming packets for a short period of time instead of filtering external traffic would further increase stealthiness, as we would not need to send artificial packets. %
This leaves the \gls{VM} with the possibility to analyze page access times to determine if the \gls{HV} is performing memory access tracking. 
However, as we only require a maximum of around three seconds for our attack, it would be difficult to distinguish between delays caused by access tracking and other activities, such as migration of the \gls{VM}.  

Our results show that the identification of the trigger page (\autoref{sec:attack:code}) and the \texttt{packet buffer} (\autoref{sec:attack:data}) work as described. 
This enables us to 
perform the \name attack
execute \name 
against different \glspl{VM} protected with \gls{SEV-ES} with a success rate of 100\,\%.

\section{Countermeasures}
\label{sec:countermeasures}

\name mainly relies on the missing integrity protection to remap the \gls{GFN} of our trigger to the \gls{SFN} of our payload. 
Therefore, assuring the integrity of the \gls{VM}'s memory would prevent remapping, and be sufficient to prevent the \name attack as well as others~\cite{hetzelt2017security, morbitzer2018severed, du2017secure, wilke2020sevurity, li2019exploiting, li2020crossline}.

To ensure the integrity of the \gls{VM}'s memory, AMD has announced \gls{SEV-SNP}~\cite{AMD2020SNP} in 
2020.
One of the main goals of \gls{SEV-SNP} is to provide integrity protection to the memory of \gls{SEV}-protected \glspl{VM}. 
\gls{SEV-SNP} achieves this by adding a \emph{\gls{RMP}} to the address translation process. 
As with \gls{SEV-ES}, the \gls{VM} translates a virtual page to a \gls{GFN}, and afterwards the \gls{HV} uses the \gls{SLAT} to translate the \gls{GFN} into the \gls{SFN}. 
In contrast to \gls{SEV-ES}, with \gls{SEV-SNP}, the \gls{SP} afterwards verifies the integrity of the translation 
using 
the \gls{RMP}. 
Each entry in the \gls{RMP} contains among others information to which \gls{VM} the \gls{SFN} belongs to, the matching \gls{GFN} and a \emph{valid} flag.
If an attacker modifies the \gls{SLAT} tables 
to point a \gls{GFN}
to a different \gls{SFN}, the \gls{SP} will be able to detect this modification and the address translation will fault with a \texttt{\#PF} exception. 
While \gls{SEV-SNP} would therefore not prevent 
determining the location of our trigger and the \texttt{packet buffer} in memory, it would prevent 
mapping the \gls{GFN} of the trigger to the \gls{SFN} of the payload.
Yet, up to this date, AMD has only published a whitepaper on \gls{SEV-SNP}~\cite{AMD2020SNP}, without announcing when \gls{SEV-SNP} will be released in hardware. 
At the same time, the currently available hardware, among others already used by cloud providers~\cite{google2021confidential}, is affected by \name, posing a threat to users relying on the confidentiality and integrity properties of \gls{SEV-ES}.

\section{Discussion}
\label{sec:discussion}

\textbf{Kernel image availability:}
In Section \autoref{sec:attack:code}, we assume 
access to the kernel image to extract the kernel's symbol information. 
This is a valid assumption as the kernel image typically resides in the unencrypted \texttt{/boot} partition of the \gls{VM}'s virtual disk image or, in some situations, can be directly provided by the HV. 
In both cases, the HV gains direct access to the kernel image.
In case the \gls{VM} stores its kernel image in an encrypted \texttt{/boot} partition, an attacker can 
use
other methods to gain access to the \gls{VM}'s kernel image. 
For example, 
\gls{SEV-ES} supports and disseminates only a small set of Linux distributions, each running a specific kernel
version 
known to the public~\cite{AMDSEV}. 
Hence,
assuming the \gls{VM}'s provider did not compile a custom Linux kernel, the attacker can 
inspect the
Linux kernel images to collect the necessary information to identify the trigger point. 
Alternatively, to determine the used kernel version---or rather the addresses to relevant kernel functions---we could apply \gls{OS} fingerprinting mechanisms~\cite{buhren2018detectability, werner2019severest}.

\smallskip 
\noindent
\textbf{KASLR:}
We have proposed one approach to practically disable \gls{KASLR}, and two approaches to identify the location 
\texttt{do\_nmi()} 
within the \gls{VM}'s memory with \gls{KASLR} 
(\autoref{sec:attack:code}).
Both identification approaches work with the most recent implementation of \gls{KASLR}. 
Yet, the granularity of \gls{KASLR} might change in the near future. 
In a recent patch, Accardi~\cite{edge2020finer} proposed to also randomize the offsets of functions within the kernel image. 
Using \gls{SEV-ES}, after injection of an interrupt, the \gls{VM} first internally handles the interrupt, which may cause it to access different \glspl{GFN} before handing control to the \gls{NMI} handler. 
To still be able to determine the location of the \gls{NMI} handler and the \texttt{do\_nmi()} page, we could analyze the page faults caused by the \gls{VM}'s internal handler to determine when it finishes execution and hands control to the \gls{NMI} handler. 
Alternatively, we could also resort to machine learning~\cite{buhren2018detectability} to determine the location of the \texttt{do\_nmi()} function.

\smallskip 
\noindent
\textbf{Injected code size:}
Our \gls{PoC} injects a small payload
executing a hypercall that allows us to determine the success of the attack. 
Further, we also experimented with payloads up to 78 bytes. 
Yet, to inflict significant damage, the attacker would prefer to execute arbitrary code. 
Thus, the injected payload size could exceed the packet size limits imposed by the \gls{MTU}.
Yet, by injecting multiple payload fragments and using \texttt{RIP}-relative jumps, we can transfer the control-flow to other payload segments 
located in different fragments. 
This way, we can inject and execute payloads of a nearly arbitrary size.

\smallskip 
\noindent
\textbf{Stealthiness:} Our \gls{PoC} is not designed to be stealthy. Yet, with some
modifications, we can reduce its visibility. 
As part of 
\name,
we remap the \gls{GFN} 
holding
the \texttt{do\_nmi()} function to the \gls{SFN} 
holding
our injected payload.
Since we can cause the \gls{VM} to execute the \texttt{do\_nmi()} function from outside of the \gls{VM}, it acts as a trigger for executing the injected payload.
To still ensure consistent \gls{VM} operation, we continue tracking the
execution of incoming \gls{NMI} requests by monitoring execute accesses
to the \gls{GFN} 
containing
the \gls{NMI} handler. 
That is, every time a regular \gls{NMI} request occurs, we can restore the
original mapping in the global \gls{SLAT} tables and continue execution.
Yet, walking the \gls{SLAT} tables to change permissions of \glspl{GFN} on
every \gls{NMI} event is 
a slow operation and 
can lead to race
conditions in multi-\gls{vCPU} environments.  Besides, changes in the \gls{VM}'s
memory are visible to other \glspl{vCPU}.  To accommodate these issues, we can
prepare a set of \gls{SLAT} tables, which define different views (i.e.,
mappings) on the \gls{VM}'s physical memory, and rapidly switch among them, without revealing the in-memory artifacts~\cite{xen-altp2m,
proskurin:2018b, xmp_proskurin:2020}. Since we can assign one \gls{SLAT} table
to each \gls{vCPU} individually, the remaining \glspl{vCPU} would not be able
to detect the adjustments. %

Further, even though our \gls{PoC} blocks external traffic to the \gls{VM} (\autoref{sec:evaluation}), we underline that this strategy is not a requirement. 
On the contrary, additional network traffic would allow us to detect the \texttt{packet buffer} in the encrypted memory. 
This way, we would not need 
artificial traffic to observe the sequences of write accesses of the bounce buffers. The same applies to identifying a suitable fragment in the \texttt{packet buffer}. 
Therefore, analyzing external traffic
increases the stealthiness of the attack; essentially, the attacker would have to send only a single packet with the payload. 
To avoid issues caused by multiple network packets received shortly after each other, an improved implementation of our \gls{PoC} can queue packets until analysis steps such as access page tracking of previous packets are finished. 
Therefore, \name can also be applied to \glspl{VM} with high network load. 
However, we opted for an evaluation 
without additional
network traffic in our test environment, as this could have influenced our evaluation results. 

Besides, in our \gls{PoC}, we send UDP packets to a \emph{closed} port. 
The \gls{VM} can detect the unusual amount of 
traffic to a closed port and reveal the attack.
To further increase the stealthiness of our approach, we could fall back to sending less suspicious packets, such as ARP packets. 
While ARP packets are limited in size, we can use them to detect the \texttt{packet buffer} and to fill unsuited \texttt{packet buffer} fragments with dummy packets.

\smallskip 
\noindent
\textbf{Generalization:}
While we discussed \name by injecting a payload via network packets, we neither rely on the usage of virtio or Linux, nor the network communication channel.
Even though \name uses Linux and virtio as concrete examples,
the actual vulnerability is entirely independent from the \gls{OS}, subsystem and driver. 
Assuming that another \gls{OS}, such as Windows, would provide support for \gls{SEV-ES}, the system would still be architecturally forced to transfer data from DMA-capable devices into the protected \gls{VM}'s memory. 
The system will achieve this 
by copying the data from unencrypted to encrypted memory, which we exploit to inject our payload.
Therefore, targeting another \gls{OS}, subsystem or driver would not change the fundamentals of the three steps performed by \name. 

However, currently 
\gls{SEV-ES} is only supported by Linux, and most Linux-based virtualization environments use virtio. 
For both Linux and virtio, our current implementation of \name does not require a specific version. 
To be precise, by monitoring virtio's \texttt{VRING\_AVAIL\_F\_NO\_INTERRUPT} flag as well as its \texttt{used\_event} index, \name can be applied to older as well as newer versions of virtio. 
Additionally, 
we tested \name on different kernel versions between \texttt{v4.19.43} and \texttt{v5.7.0}. 
The kernel version is only limited by having to support \gls{SEV}, which has been integrated into the Linux mainline kernel \texttt{v4.16}~\cite{AMDSEV}.

\section{Related Work}
\label{sec:related}

Hetzelt and Buhren performed the first in-depth analysis of \gls{SEV}~\cite{hetzelt2017security}.
They made use of the lacking integrity protection and combined it with the \gls{HV}'s ability to modify the \gls{SLAT} tables. 
By analyzing syscall sequences, the authors were able to modify the \gls{VM}'s login information for an SSH server,
gaining unauthenticated access to the \gls{VM}. 
In comparison to \name, this attack requires interaction of the victim, which will happen only rarely on production systems. 
Another difference is that 
the attack requires an in-depth analysis of the target, for example by having access to a comparable \gls{VM}. 
To perform such an analysis, the same authors also discussed to use machine learning to understand the behavior of the \gls{VM}, which also requires a significant amount of preparation~\cite{buhren2018detectability}.

Morbitzer et al.~\cite{morbitzer2018severed} showed how a malicious \gls{HV} can extract plaintext memory content from an encrypted \gls{VM}. 
The attack 
uses
a service in the \gls{VM} to extract the content of arbitrary memory pages after modifying the \gls{SLAT} tables. 
Compared to \name, this approach requires a high number of requests. 
Therefore, Morbitzer et al.~\cite{morbitzer2019extracting} improved the attack 
to perform targeted extraction of secrets from a \gls{VM}.
This drastically reduced the 
memory pages
to be extracted, 
therefore also reducing
the time required for the attack. 
A disadvantage that remains is that the target \gls{VM} is required to provide an in-\gls{VM} service,
such as a web server. 

Du et al.~\cite{du2017secure} analyzed the encryption algorithm of \gls{SEV}.
They determined that before encryption, the plaintext is tweaked with static values depending on the System Physical Address on which the data will be stored. 
This knowledge enabled them to gain access to an SSH server running within the encrypted \gls{VM} by analyzing cipher blocks. 
The known tweak values were also used by Li et al.~\cite{li2019exploiting}, who built a decryption oracle by replacing encrypted memory blocks about to be sent via an SSH connection with arbitrary blocks from the same \gls{VM}. 
Using a similar technique, they were also able to create an encryption oracle. %
Building on the known tweak values, Wilke et al.~\cite{wilke2020sevurity} moved cipher blocks in the \gls{VM}'s encrypted memory, allowing them to execute code gadgets within the \gls{VM}. 
However, all three attacks rely on knowing the tweak values 
to be able to meaningfully replace encrypted memory blocks in the \gls{VM}'s memory. 
While these tweak values were static on the first generation of EPYC CPUs, the second generation randomly generates the tweak values at every boot~\cite{wilke2020sevurity}, making it resistant against the attacks. 
In comparison, \name does not rely on knowing the tweak mechanism of the memory encryption to execute arbitrary code. 

Werner et al.~\cite{werner2019severest} monitored register values and memory accesses of an encrypted \gls{VM} to derive and modify instructions the \gls{VM} executed. 
Further, they used \gls{IBS} to gather information about which applications were running within the encrypted \gls{VM}. 
Compared to their approach to read and modify confidential information via register values, their application fingerprinting technique can also be applied to \gls{SEV-ES}. 
Therefore, it could be used as a base to implement \gls{OS} fingerprinting to avoid the need to access the \gls{VM} kernel's symbol information 
to determine 
the trigger function's page offset
(\autoref{sec:attack:code}).

Li et al.~\cite{li2020crossline} presented 
an attack
in which the \gls{HV} modifies the Address Space Identifier of a \gls{VM} during a \texttt{VMEXIT}, 
allowing
them to decrypt the page tables of the target \gls{VM}. 
By additionally modifying the \gls{VM}'s state, they 
created 
decryption and encryption oracles.
As this modification of a \gls{VM}'s state is prevented by \gls{SEV-ES}, their encryption and decryption oracles cannot be applied to \glspl{VM} protected by \gls{SEV-ES}. 
In comparison, \name can be applied to \gls{SEV} as well as to \gls{SEV-ES}-protected \glspl{VM}.

Radev and Morbitzer~\cite{radev2020abusing} 
manipulated the interfaces of a \gls{VM} protected by \gls{SEV-ES}. 
This allowed them to trick the \gls{VM} into exposing confidential data to the \gls{HV}, and to reduce the entropy of the \gls{VM}'s kernel probabilistic defenses, 
Additionally, they managed to execute code within the \gls{VM} by providing false information about \gls{SEV} support and adding additional MMIO regions.  
All of the proposed attacks make use of software bugs which have meanwhile been patched~\cite{radev2020patch, roedel2020patch}. 
In comparison, \name 
uses
a fundamental design flaw in \gls{SEV-ES}, which will require hardware changes to be fully prevented. 

\section{Conclusion}

In this work, we have presented \name, a powerful attack against AMD
\gls{SEV-ES}, which, compared to previous work, allows to inject and execute
\emph{truly arbitrary} code into encrypted \glspl{VM}.
The attack builds upon \gls{SEV-ES}'s lack of memory integrity protection and its
inability to perform \gls{DMA} into encrypted memory.  
Compared to previous work, \name 
allows to inject and execute arbitrary code into protected \glspl{VM} that cannot be prevented by currently available countermeasures. 
At the same time, we do not require any open ports in the \gls{VM}, as we abuse the fact that Ethernet frames are pulled into the \gls{VM}'s memory before any filtering mechanisms can apply.
Also, it suffices if our packets are copied into the \gls{VM}'s memory only for a short period of time. 

We evaluated 
\name 
against various \glspl{VM} protected with \gls{SEV-ES}. 
Our evaluation demonstrates a success rate of 100\,\%, for which we need only a few seconds. %
Overall, our work is the next step in a series of attacks which highlight that AMD 
\gls{SEV-ES} will not be able to provide sufficient security guarantees without protecting the integrity of the encrypted memory.  
Yet, we are optimistic that AMD's next evolution, \gls{SEV-SNP}, will eliminate the weaknesses of its predecessors, providing stronger security guarantees to \glspl{VM} in untrusted environments.

\section*{Acknowledgment}

This work has been partially funded by the Fraunhofer Cluster of Excellence ``Cognitive Internet Technologies''. %

\bibliographystyle{IEEEtran}
\bibliography{biblio}

\begin{thebibliography}{10}
\providecommand{\url}[1]{#1}
\csname url@samestyle\endcsname
\providecommand{\newblock}{\relax}
\providecommand{\bibinfo}[2]{#2}
\providecommand{\BIBentrySTDinterwordspacing}{\spaceskip=0pt\relax}
\providecommand{\BIBentryALTinterwordstretchfactor}{4}
\providecommand{\BIBentryALTinterwordspacing}{\spaceskip=\fontdimen2\font plus
\BIBentryALTinterwordstretchfactor\fontdimen3\font minus
  \fontdimen4\font\relax}
\providecommand{\BIBforeignlanguage}[2]{{%
\expandafter\ifx\csname l@#1\endcsname\relax
\typeout{** WARNING: IEEEtran.bst: No hyphenation pattern has been}%
\typeout{** loaded for the language `#1'. Using the pattern for}%
\typeout{** the default language instead.}%
\else
\language=\csname l@#1\endcsname
\fi
#2}}
\providecommand{\BIBdecl}{\relax}
\BIBdecl

\bibitem{denisco2019cloud}
A.~DeNisco~Rayome, ``{69\% of enterprises moving business-critical applications
  to the cloud},''
  \url{https://www.techrepublic.com/article/69-of-enterprises-moving-business-critical-applications-to-the-cloud/},
  January 2019, {Accessed: 2021-27-01}.

\bibitem{amigorena2019why}
F.~Amigorena, ``{Why SMBs Still do not Trust Cloud Storage Providers to Secure
  their Data},''
  \url{https://www.infosecurity-magazine.com/opinions/smb-trust-cloud-storage-1-1/},
  2019, {Accessed: 2021-27-01}.

\bibitem{liu2015thwarting}
Y.~Liu, T.~Zhou, K.~Chen, H.~Chen, and Y.~Xia, ``{Thwarting Memory Disclosure
  with Efficient Hypervisor-Enforced Intra-Domain Isolation},'' in \emph{ACM
  Conference on Computer and Communications Security (CCS)}, 2015.

\bibitem{chen:2017:privwatcher}
Q.~Chen, A.~M. Azab, G.~Ganesh, and P.~Ning, ``{PrivWatcher: Non-bypassable
  Monitoring and Protection of Process Credentials from Memory Corruption
  Attacks},'' in \emph{ACM Symposium on Information, Computer and
  Communications Security (ASIACCS)}, 2017.

\bibitem{hua2018epti}
Z.~Hua, D.~Du, Y.~Xia, H.~Chen, and B.~Zang, ``{EPTI: Efficient Defence against
  Meltdown Attack for Unpatched VMs},'' in \emph{USENIX Annual Technical
  Conference}, 2018.

\bibitem{xmp_proskurin:2020}
S.~Proskurin, M.~Momeu, S.~Ghavamnia, V.~P. Kemerlis, and M.~Polychronakis,
  ``{xMP: Selective Memory Protection for Kernel and User Space},'' in
  \emph{IEEE Symposium on Security and Privacy (S\&P)}, 2020.

\bibitem{vmware2017vmsa}
{VMware Inc.}, ``{VMSA-2017-0018.1},''
  \url{https://www.vmware.com/security/advisories/VMSA-2017-0018.html},
  November 2017, {Accessed: 2021-27-01}.

\bibitem{ssd2018oracle}
{SSD Secure Disclosure}, ``{SSD Advisory – Oracle VirtualBox Multiple Guest
  to Host Escape Vulnerabilities},''
  \url{https://ssd-disclosure.com/ssd-advisory-oracle-virtualbox-multiple-guest-to-host-escape-vulnerabilities/},
  January 2018, {Accessed: 2021-27-01}.

\bibitem{citrix2019hypervisor}
{Citrix Systems}, ``{Citrix Hypervisor Security Update},''
  \url{https://support.citrix.com/article/CTX263477}, 2019, {Accessed:
  2021-27-01}.

\bibitem{citrix2020hypervisor}
------, ``{Citrix Hypervisor Security Update},''
  \url{https://support.citrix.com/article/CTX286756}, 2020, {Accessed:
  2021-27-01}.

\bibitem{becher2005firewire}
M.~Becher, M.~Dornseif, and C.~N. Klein, ``Firewire: all your memory are belong
  to us,'' \emph{Proceedings of CanSecWest}, 2005.

\bibitem{boileau2006hit}
A.~Boileau, ``Hit by a bus: Physical access attacks with firewire,''
  \emph{Presentation, Ruxcon}, vol.~3, 2006.

\bibitem{halderman2008lest}
J.~A. Halderman, S.~D. Schoen, N.~Heninger, W.~Clarkson, W.~Paul, J.~A.
  Calandrino, A.~J. Feldman, J.~Appelbaum, and E.~W. Felten, ``{Lest We
  Remember: Cold Boot Attacks on Encryption Keys},'' in \emph{USENIX Security
  Symposium}, 2008.

\bibitem{lengyel:2014}
T.~K. Lengyel, S.~Maresca, B.~D. Payne, G.~D. Webster, S.~Vogl, and A.~Kiayias,
  ``{Scalability, Fidelity and Stealth in the DRAKVUF Dynamic Malware Analysis
  System},'' in \emph{Annual Computer Security Applications Conference
  (ACSAC)}, 2014.

\bibitem{proskurin:2018a}
S.~Proskurin, J.~Kirsch, and A.~Zarras, ``{Follow the WhiteRabbit: Towards
  Consolidation of On-the-Fly Virtualization and Virtual Machine
  Introspection},'' in \emph{IFIP International Conference on ICT Systems
  Security and Privacy Protection (IFIP SEC)}, 2018.

\bibitem{proskurin:2018b}
S.~Proskurin, T.~Lengyel, M.~Momeu, C.~Eckert, and A.~Zarras, ``{Hiding in the
  Shadows: Empowering ARM for Stealthy Virtual Machine Introspection},'' in
  \emph{Annual Computer Security Applications Conference (ACSAC)}, 2018.

\bibitem{meixner2012trust}
F.~Meixner and R.~Buettner, ``Trust as an integral part for success of cloud
  computing,'' in \emph{International Conference on Internet and Web
  Applications and Services (ICIW)}, 2012.

\bibitem{kaplan2016amd}
D.~Kaplan, J.~Powell, and T.~Woller, ``{AMD Memory Encryption},'' Advanced
  Micro Devices, Tech. Rep., 2016.

\bibitem{singh2017x86}
B.~Singh, ``{x86: Secure Encrypted Virtualization (AMD)},''
  \url{https://lwn.net/Articles/716165/}, March 2017, {Accessed: 2021-27-01}.

\bibitem{morbitzer2018severed}
M.~Morbitzer, M.~Huber, J.~Horsch, and S.~Wessel, ``{SEVered: Subverting AMD's
  Virtual Machine Encryption},'' in \emph{European Workshop on Systems Security
  (EuroSEC)}, 2018.

\bibitem{morbitzer2019extracting}
M.~Morbitzer, M.~Huber, and J.~Horsch, ``{Extracting Secrets from Encrypted
  Virtual Machines},'' in \emph{ACM Conference on Data and Application Security
  and Privacy (CODASPY)}, 2019.

\bibitem{werner2019severest}
J.~Werner, J.~Mason, M.~Antonakakis, M.~Polychronakis, and F.~Monrose, ``{The
  SEVerESt Of Them All: Inference Attacks Against Secure Virtual Enclaves},''
  in \emph{ACM Symposium on Information, Computer and Communications Security
  (ASIACCS)}, 2019.

\bibitem{li2019exploiting}
M.~Li, Y.~Zhang, Z.~Lin, and Y.~Solihin, ``{Exploiting Unprotected I/O
  Operations in AMD{\textquoteright}s Secure Encrypted Virtualization},'' in
  \emph{USENIX Security Symposium}, 2019.

\bibitem{hetzelt2017security}
F.~Hetzelt and R.~Buhren, ``{Security Analysis of Encrypted Virtual
  Machines},'' in \emph{International Conference on Virtual Execution
  Environments}, 2017.

\bibitem{du2017secure}
\BIBentryALTinterwordspacing
Z.-H. Du, Z.~Ying, Z.~Ma, Y.~Mai, P.~Wang, J.~Liu, and J.~Fang, ``{Secure
  Encrypted Virtualization is Unsecure},'' 2017. [Online]. Available:
  \url{https://arxiv.org/abs/1712.05090}
\BIBentrySTDinterwordspacing

\bibitem{wilke2020sevurity}
L.~Wilke, J.~Wichelmann, M.~Morbitzer, and T.~Eisenbarth, ``{SEVurity: No
  Security Without Integrity - Breaking Integrity-Free Memory Encryption with
  Minimal Assumptions},'' in \emph{IEEE Symposium on Security and Privacy
  (S\&P)}, 2020.

\bibitem{radev2020abusing}
M.~Radev and M.~Morbitzer, ``{Exploiting Interfaces of Secure Encrypted Virtual
  Machines},'' in \emph{Reversing and Offensive-oriented Trends Symposium},
  2020.

\bibitem{cohen2019sev}
{Cfir Cohen}, ``{AMD-SEV: Platform DH key recovery via invalid curve attack:
  (CVE-2019-9836)},'' \url{https://seclists.org/fulldisclosure/2019/Jun/46},
  June 2019, {Accessed: 2021-27-01}.

\bibitem{buhren2019insecure}
R.~Buhren, C.~Werling, and J.-P. Seifert, ``{Insecure Until Proven Updated:
  Analyzing AMD SEV's Remote Attestation},'' in \emph{ACM Conference on
  Computer and Communications Security (CCS)}, 2019.

\bibitem{sev-es}
{David Kaplan}, ``{Protecting VM Register State with SEV-ES},'' White Paper,
  2017.

\bibitem{buhren2018detectability}
R.~Buhren, F.~Hetzelt, and N.~Pirnay, ``On the detectability of control flow
  using memory access patterns,'' in \emph{Workshop on System Software for
  Trusted Execution (SysTEX)}, 2018.

\bibitem{oasis2019virtio}
{Edited by Michael S. Tsirkin and Cornelia Huck}, \emph{{Virtual I/O Device
  (VIRTIO) Version 1.1}}, {OASIS Committee Specification 01}, 2019, latest
  version:
  \url{https://docs.oasis-open.org/virtio/virtio/v1.1/virtio-v1.1.html.}

\bibitem{tang2012cleanos}
Y.~Tang, P.~Ames, S.~Bhamidipati, A.~Bijlani, R.~Geambasu, and N.~Sarda,
  ``{CleanOS: Limiting Mobile Data Exposure with Idle Eviction},'' in
  \emph{USENIX Symposium on Operating System Design and Implementation (OSDI)},
  2012.

\bibitem{weinmann2012baseband}
R.-P. Weinmann, ``{Baseband Attacks: Remote Exploitation of Memory Corruptions
  in Cellular Protocol Stacks},'' in \emph{USENIX Workshop on Offensive
  Technologies (WOOT)}, 2012.

\bibitem{iommu_ben:2006}
M.~Ben-Yehuda, J.~Mason, J.~Xenidis, O.~Krieger, L.~Van~Doorn, J.~Nakajima,
  A.~Mallick, and E.~Wahlig, ``{Utilizing IOMMUs for Virtualization in Linux
  and Xen},'' in \emph{Ottawa Linux Symposium}, 2006.

\bibitem{AMD2020}
{Advanced Micro Devices}, ``{AMD64 Architecture Programmer’s Manual (Volumes
  1-5)},'' 2020.

\bibitem{kaslr_edge:2013}
J.~Edge, ``{Kernel Address Space Layout Randomization},''
  \url{https://lwn.net/Articles/569635/}, 2013, {Accessed: 2021-27-01}.

\bibitem{kernel2020self}
{The kernel development community}, ``{Kernel Self-Protection},''
  \url{https://www.kernel.org/doc/html/latest/_sources/security/self-protection.rst.txt},
  2020, {Accessed: 2021-27-01}.

\bibitem{schwarz2019store}
\BIBentryALTinterwordspacing
M.~Schwarz, C.~Canella, L.~Giner, and D.~Gruss, ``{Store-to-Leak
  Forwarding:Leaking Data on Meltdown-resistant CPUs},'' 2019. [Online].
  Available: \url{https://arxiv.org/abs/1905.05725}
\BIBentrySTDinterwordspacing

\bibitem{dalton2013mergeable}
M.~Dalton, ``{[PATCH net-next 4/4] virtio-net: auto-tune mergeable rx buffer
  size for improved performance},''
  \url{https://lists.linuxfoundation.org/pipermail/virtualization/2013-November/025626.html},
  2013, {Accessed: 2021-27-01}.

\bibitem{ieee2018ethernet}
``{IEEE Standard for Ethernet},'' \emph{IEEE Std 802.3-2018 (Revision of IEEE
  Std 802.3-2015)}, pp. 1--5600, 2018.

\bibitem{severedframework}
M.~Morbitzer and M.~Huber, ``{Github - SEVered Framework},''
  \url{https://github.com/Fraunhofer-AISEC/severed-framework/}, 2019.

\bibitem{AMDSEV}
{Advanced Micro Devices}, ``{GitHub - AMDESE/AMDSEV: AMD Secure Encrypted
  Virtualization},'' \url{https://github.com/AMDESE/AMDSEV}, 2018, {Accessed:
  2021-27-01}.

\bibitem{li2020crossline}
\BIBentryALTinterwordspacing
M.~Li, Y.~Zhang, and Z.~Lin, ``{CROSSLINE: Breaking "Security-by-Crash" based
  Memory Isolation in AMD SEV},'' 2020. [Online]. Available:
  \url{https://arxiv.org/abs/2008.00146}
\BIBentrySTDinterwordspacing

\bibitem{radev2020patch}
M.~Radev, ``{[PATCH v7 72/72] x86/sev-es: Check required CPU features for
  SEV-ES},'' \url{https://lkml.org/lkml/2020/9/7/1074}, 2020, {Accessed:
  2021-27-01}.

\bibitem{roedel2020patch}
J.~Roedel, ``{[PATCH v2 0/5] x86/sev-es: Mitigate some HV attack vectors},''
  \url{https://lkml.org/lkml/2020/10/20/465}, 2020, {Accessed: 2021-27-01}.

\bibitem{AMD2020SNP}
{Advanced Micro Devices}, ``{AMD SEV-SNP: Strengthening VM Isolation with
  Integrity Protection and More},''
  \url{https://www.amd.com/system/files/TechDocs/SEV-SNP-strengthening-vm-isolation-with-integrity-protection-and-more.pdf},
  2020.

\bibitem{google2021confidential}
{Google}, ``Confidential vm and compute engine,'' 2021,
  \url{https://cloud.google.com/compute/confidential-vm/docs/about-cvm}.

\bibitem{edge2020finer}
J.~Edge, ``{Finer-grained kernel address-space layout randomization},''
  \url{https://lwn.net/Articles/812438/}, 2020, {Accessed: 2021-27-01}.

\bibitem{xen-altp2m}
{Xen-devel mailing list}, ``{Alternate \texttt{p2m} design specification},''
  \url{https://lists.xenproject.org/archives/html/xen-devel/2015-06/msg01319.html},
  2015, {Accessed: 2021-27-01}.

\end{thebibliography}

\end{document}